\documentclass{mc-abstract}

\year{2023}

\title{Quantum Microservices Development and Deployment}
\author{Enrique Moguel$^1$ \and Jose Garcia-Alonso$^2$ \and Majid Haghparast$^3$ \and Juan M. Murillo $^1$}
\affiliation{
    $^1$ Computing and Advanced Technologies Foundation of Extremadura, Carretera Nacional 521, Km 41,8, Cáceres, 10071, Spain \\ 
    $^2$ Quercus Software Engineering Group, Universidad de Extremadura, Avda. de la Universidad, s/n, Cáceres, 10004, Spain \\
    $^3$ Faculty of Information Technology, University of Jyväskylä, 40014 Jyväskylä, Finland
}
\correspondence{enrique@unex.es}

\begin{document}

\maketitle

\abstract{Early advances in the field of quantum computing have provided new opportunities to tackle intricate problems in areas as diverse as mathematics, physics, or healthcare. However, the technology required to construct such systems where different pieces of quantum and classical software collaborate is currently lacking. For this reason, significant advancements in quantum service-oriented computing are necessary to enable developers to create and operate quantum services and microservices comparable to their classical counterparts. Therefore, the core objective of this work is to establish the necessary technological infrastructure that enables the application of the benefits and lessons learned from service-oriented computing to the domain of quantum software engineering. To this end, we propose a pipeline for the continuous deployment of services. Additionally, we have validated the proposal by making use of a modification of the OpenAPI specification, the GitHub Actions, and AWS.}

%%%%%%%%%%%%%%%%%%%%%%%%%%%%%%%%%%%%%%%%%%%%%%%%%%
%%%%%%%%%%%%%%%%%%%%%%%%%%%%%%%%%%%%%%%%%%%%%%%%%%
\section{Introduction}

Quantum computing has garnered considerable attention across a multitude of research domains, encompassing mathematics and physics, with a primary emphasis on constructing increasingly potent quantum computers and enhancing qubit stability. Additionally, disciplines such as economics and healthcare have focused on formulating novel quantum algorithms tailored to specific problem domains \cite{Preskill2018}. Nevertheless, the current research landscape exhibits a dearth of concerted efforts toward the advancement of technologies dedicated to the construction of quantum software \cite{Piattini2020}.

Contemporary information systems exhibit a pervasive complexity arising from their composition of discrete components that are globally distributed and interconnected through intricate communication infrastructures and protocols. These components, which tend to be encapsulated as services, possess well-defined responsibilities. The realization of such systems is made feasible through a range of technologies, including Service Oriented Computing and cloud computing \cite{Papazoglou2003}. Adopting this architectural approach to software system development confers several advantages, such as cost optimization by paying solely for utilized infrastructure, and the attainment of desirable attributes such as heightened interoperability, autonomy, reduced coupling, reusability, maintainability, reliability, scalability, and security.

% Presently, the development of quantum computers is underway to support the execution of quantum software. While the future of quantum software remains uncertain, certain features can be confidently predicted. Firstly, quantum software is expected to coexist alongside classical software and information systems. Secondly, its primary focus will be on solving problems that are beyond the capabilities of classical systems. As a result, collaboration between classical and quantum systems, as well as their respective software components, will be essential. Thirdly, the utilization of service-oriented principles from the software engineering discipline is widely regarded as the most effective approach for managing the collaboration between heterogeneous systems.

While the future of quantum software remains uncertain, certain features can be confidently predicted: 1) quantum software is expected to coexist alongside classical software and information systems; 2) its primary focus will be on solving problems that are beyond the capabilities of classical systems; and 3) the utilization of service-oriented principles is widely regarded as the most effective approach for managing the collaboration between heterogeneous systems.

However, the technology required to construct such systems where different pieces of quantum and classical software collaborate is currently lacking \cite{Moguel2022}. For this reason, significant advancements in quantum service-oriented computing are necessary to enable developers to create and operate quantum services and microservices comparable to their classical counterparts. These advancements will pave the way for the creation of hybrid systems that seamlessly integrate quantum and classical components, offering a unified and cohesive environment for collaborative problem-solving.

In summary, Quantum Software Engineering is expected to become a prominent field motivating software engineers. To realize this vision, the development of techniques and tools that streamline the construction of complex systems within the realm of Quantum Computing is imperative. 
%This objective has garnered recognition from esteemed institutions such as the Software Engineering Institute at Carnegie Mellon University, evident in their publication titled "A National Agenda for Software Engineering Research \& Development: Architecting the Future of Software Engineering." This agenda highlights the significance of Quantum Computing Software Systems as a focal point, emphasizing the societal value and potential benefits derived from quantum software.

Therefore, the core objective of this work is to establish the necessary technological infrastructure that enables the application of the benefits and lessons learned from service-oriented computing to the domain of quantum software engineering. Additionally, we have validated the proposal by making use of a modification of the OpenAPI specification, the GitHub Actions, and AWS.

%%%%%%%%%%%%%%%%%%%%%%%%%%%%%%%%%%%%%%%%%%%%%%%%%%
%%%%%%%%%%%%%%%%%%%%%%%%%%%%%%%%%%%%%%%%%%%%%%%%%%
\section{Background}

The power of quantum computers stems from their ability to solve problems categorized as bounded-error quantum polynomial time (BQP) problems. These are decision problems that can be solved by a quantum computer in polynomial time, albeit with a certain margin of error. It is currently conjectured that the class BQP encompasses class P, signifying that quantum computers can solve problems that classical computers can also solve efficiently. However, BQP is also believed to include problems that lie outside the class P, making them of great interest to researchers.

% There are two key aspects regarding these problems. Firstly, they are instrumental in demonstrating the concept of quantum supremacy, which asserts that a programmable quantum device can solve a problem that is infeasible for any classical computer to solve within a practical timeframe. Claims of achieving quantum supremacy have been put forth by Google \cite{Arute2019} and other research teams \cite{Wu2021}. These demonstrations serve as compelling evidence of the potential superiority of quantum computing over classical computing.

% Secondly, these problems have practical applications, such as prime factorization or the simulation of quantum systems. Prime factorization is of particular significance due to its relevance in cryptography, where the ability to efficiently factor large numbers is crucial. Furthermore, the simulation of quantum systems is valuable for studying and understanding complex quantum phenomena, which can have implications in various scientific and technological domains.

In conclusion, the exploration and utilization of problems within the BQP class hold both theoretical and practical significance, ranging from showcasing quantum supremacy to enabling advancements in areas like cryptography and quantum system simulation \cite{Arute2019}.

Indeed, the coexistence of classical and quantum computers is a natural consequence of the current state of quantum technology, which is still expensive and in its early stages of development. The transition to the new quantum computing paradigm cannot happen instantaneously, necessitating the parallel operation of classical and quantum systems. In this context, relying solely on the progress of quantum computer science is insufficient to effectively realize the potential of quantum computing and overcome the existing challenges.

To address this need, the establishment of a new field called "Quantum Software Engineering" is imperative. This field focuses on innovating software development methodologies that effectively map practical problems onto quantum computers \cite{Maslov2019}. The goal is to enable quantum software engineers to design and implement impactful software applications that leverage the computational speed and capabilities offered by quantum computers \cite{Mueck2017}. Various aspects of software engineering, ranging from requirement analysis to software reuse, require reevaluation and adaptation to accommodate the development of complex quantum software \cite{Piattini2020, Moguel2020}.

While there are numerous quantum programming languages, SDKs, and platforms available, they alone are not sufficient to meet the demands of quantum software development \cite{Piattini2020}. This is primarily due to the lack of a solid methodological foundation in quantum software engineering.

In addition, in the context of quantum Continuous Deployment (CD), there have been proposals to incorporate quantum algorithms into different stages of the CD cycle. These proposals take into account characteristics specific to quantum computing, such as the number of qubits, cloud accessibility, and error rates of quantum machines. However, no tools or implementations have been developed to assist developers in this process. As a result, despite the existence of proposals to integrate quantum algorithms into CD stages, certain aspects of quantum CD remain unaddressed. Specifically, critical phases like design and deployment require careful consideration. % These phases involve the creation of the software package and its deployment into the production environment. 
This paper aims to cover these aspects and optimize the generation and deployment processes, thereby accelerating the delivery of quantum computing applications.

Considering the limitations and the anticipated migration towards hybrid systems where classical software interfaces with quantum algorithms \cite{PerezCastillo2021}, this paper specifically focuses on Service Engineering for Quantum systems. By applying service engineering principles to quantum systems, the aim is to develop methodologies and techniques that facilitate the construction and integration of classical and quantum software components within hybrid systems.

%%%%%%%%%%%%%%%%%%%%%%%%%%%%%%%%%%%%%%%%%%%%%%%%%%
%%%%%%%%%%%%%%%%%%%%%%%%%%%%%%%%%%%%%%%%%%%%%%%%%%
\section{Proposal for a quantum service-oriented architecture}

\subsection{Quantum Service Engineering}

In the current landscape, most existing quantum computers are accessible through the cloud via a model known as Quantum Computing as a Service (QCaaS) \cite{Rahaman2015}. QCaaS enables developers to access quantum hardware, but it heavily relies on specific hardware configurations, requiring a high level of proficiency in quantum computing to fully harness its advantages.

To enhance the abstraction level of QCaaS, several ongoing research efforts are underway. Commercial platforms such as Amazon Braket provide a development environment for software engineers to build quantum algorithms, test them on simulators, and execute them on various quantum hardware platforms transparently. Another example is QPath, an ecosystem that integrates the classical and quantum domains within a quantum development and application lifecycle platform, catering to a wide range of potential applications with high-quality quantum software.

In the academic realm, some works have emerged in the field of quantum software engineering \cite{Zhao2020, Piattini2020b}. These works primarily focus on translating software engineering principles to the domain of quantum software. However, there are relatively few works that specifically address service engineering for quantum software.

Introducing service engineering tools to quantum software development could significantly advance the field of quantum software engineering. Emerging works, such as \cite{Piattini2020b}, propose concepts like Quantum Application as a Service (QaaS) to bridge the gap between service engineering and quantum software. These efforts aim to apply service-oriented principles to quantum software, enabling advancements in the servitization of quantum algorithms and their lifecycle management, including deployment, execution, orchestration, and more. Works like \cite{Wild2020} explore challenges related to the orchestration of quantum services, which is crucial in this context.

Moreover, alongside the development, execution, and orchestration of quantum services, it is essential not to overlook other aspects of quality service engineering \cite{Shanshan2021}. This includes aspects related to the quality assurance and testing of quantum services, as well as the security of quantum services. These considerations play a vital role in ensuring the reliability, efficiency, and safety of quantum software systems.

%%%%%%%%%%%%%%%%%%%%%%%%%%%%%%%%%%%%%%%%%%%%%%%%%%
\subsection{Development and Deployment for Quantum Software}

DevOps is a software development methodology that focuses on the integration and synchronization of development and operations teams, with the goal of optimizing the software delivery process. It emphasizes collaboration, automation, and effective communication to improve the speed, quality, and reliability of software deployment.

In the context of quantum software development, DevOps plays a crucial role in addressing the unique challenges posed by quantum computing. 
%Quantum software requires rigorous testing and validation procedures due to quantum noise-induced errors. By employing DevOps practices, these testing and validation processes can be automated, leading to more dependable quantum software. Additionally, quantum software often exhibits higher complexity and computational requirements compared to classical software. DevOps can assist in managing this complexity by automating various aspects of quantum software development, such as deployment, scaling, and maintenance. This automation improves the efficiency and scalability of quantum software throughout its lifecycle.

Continuous Integration (CI) and Continuous Deployment (CD) are essential practices in the DevOps methodology that aim to automate the software delivery process. CI involves integrating code changes into a shared repository, while CD focuses on automating the deployment of software updates to the production environment as soon as they are ready. Efforts are being made to automate these processes using available commercial tools.

\begin{figure}
    \centering
    \includegraphics[width=0.8\textwidth]{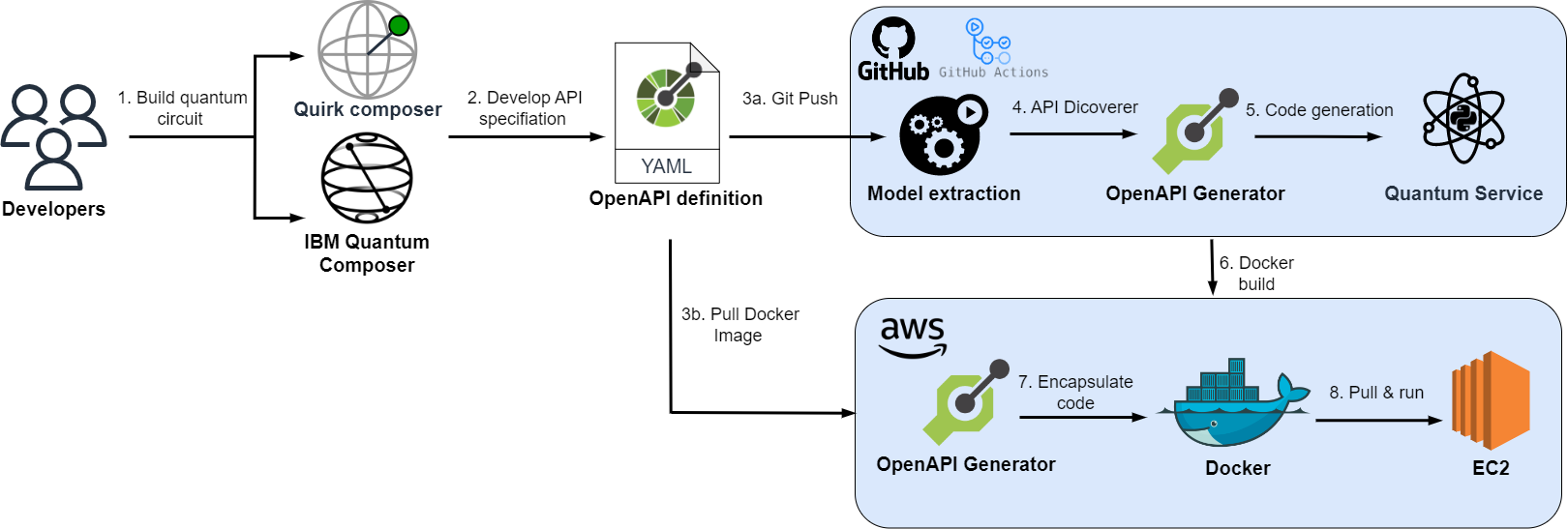}
    \caption{Continuous Deployment for Quantum Software}
    \label{fig:CD}
\end{figure}

For all these reasons, we propose the continuous deployment architecture shown in Figure \ref{fig:CD}. The first step is to define the business logic of the service as a quantum circuit using Open Quirk (indicating the Open Quirk URL of the created circuit) or directly indicating a URL where the source code is in Qiskit language (obtained from the IBM Quantum Composer). 

To integrate the business logic of quantum services, we utilize a graphical quantum programming tool that enables drag-and-drop operations for building quantum circuits. One such tool is Open Quirk, a quantum circuit composer developed using the JavaScript programming language. Open Quirk is an open-source software solution designed to facilitate the rapid prototyping of quantum circuits. It's important to note that while Open Quirk is the suggested quantum circuit composer, it can be easily substituted with other tools that offer similar capabilities for quantum circuit creation. The key requirement is that the chosen tool should allow programmatic access to the circuit code, enabling seamless integration with the OpenAPI Specification for the quantum service.

In the second step, the quantum API is defined by establishing an API contract using the OpenAPI Specification. This contract consists of defining multiple endpoints that correspond to different API methods, with each endpoint having its own specific business logic. To link the defined quantum circuit with a particular endpoint that will access its corresponding business logic, it is necessary to include either the Open Quirk URL of the circuit or the URL where the Qiskit code is located in the YAML specification of the API. The API definition is a YAML file that follows the standard structure of the OpenAPI Specification. This file provides comprehensive information about the API, including general details, available endpoints, paths, and the supported operations for each path. 

To facilitate Continuous Deployment, a pipeline has been developed to automate the code generation and deployment process, ensuring that the code is readily available for consumption by users. This pipeline leverages GitHub Actions, a popular tool provided by GitHub for code management. GitHub Actions enables developers to define actions that need to be performed whenever a change is detected in the repository. 

The pipeline, powered by GitHub Actions, further automates the process by generating the code for the quantum services. The manual process of the developers ends with a \textit{commit} to the repository, and the process of automatic generation and deployment of the services begins (step 3). In this step, the specification in the repository is validated to ensure proper formatting. The modified version of the OpenAPI Code Generator is employed to generate the code for the services (Steps 4 and 5). If the code is generated successfully, the subsequent task involves deploying the services in a container (Step 6). To achieve this, a request is sent to the Deployment API hosted on the AWS server. This request includes the URL pointing to the YAML file containing the specification and the necessary credentials to configure the execution on the service providers. The Deployment API then handles the deployment process, ensuring that the quantum services are made available for consumption.

By automating these steps, developers are relieved from the manual tasks of code generation and deployment. The pipeline, powered by GitHub Actions and integrated with the Deployment API, streamlines the process and enables efficient deployment of the quantum services in containers.

The server receives the call from the GitHub Actions generates and encapsulates the code in a container (step 7) and deploys it by exposing it on the first free port (step 8) at EC2 service. Once it is ready, it returns the URL where the generated services are hosted.

%%%%%%%%%%%%%%%%%%%%%%%%%%%%%%%%%%%%%%%%%%%%%%%%%%
%%%%%%%%%%%%%%%%%%%%%%%%%%%%%%%%%%%%%%%%%%%%%%%%%%
\section{Conclusion}

In conclusion, the main purpose of the work presented in this paper is to furnish a repertoire of techniques and methodologies for the advancement of quantum software development, drawing inspiration from the insights gleaned from classical software engineering. By leveraging the knowledge and experiences derived from traditional software engineering practices, this endeavor aims to enhance the efficacy and reliability of quantum software construction. By employing established methodologies, such as requirements analysis, design patterns, and testing strategies, this paper seeks to bridge the gap between classical and quantum software engineering, facilitating the creation of robust and dependable quantum software solutions.

\section*{Acknowledgement}
This work has been partially funded by MCIN/AEI/10.13039/501100011033 and by the EU “Next GenerationEU /PRTR”, by the Ministry of Science, Innovation and Universities (PID2021-1240454OB-C31,TED2021-130913B-I00,PDC2022-133465-I00). It is also supported by the QSALUD project (EXP 00135977/MIG-20201059) in the lines of action of the CDTI; by the Ministry of Economic Affairs and Digital Transformation of the Spanish Government through the Quantum ENIA - Quantum Spain project; by the EU through the Recovery, Transformation, and Resilience Plan – NextGenerationEU within the framework of the Digital Spain 2025 Agenda; by the Regional Ministry of Economy, Science and Digital Agenda (GR21133); and by the Academy of Finland (Project DEQSE 349945) and Business Finland (Project TORQS 8582/31/2022).

\bibliographystyle{plain}
\bibliography{biblio}

\end{document}